\documentclass[aps, pra, 10pt, twocolumn, superscriptaddress, floatfix,longbibliography,nofootinbib]{revtex4-1}
\usepackage[T1]{fontenc}
\usepackage[utf8]{inputenc}
\usepackage[english]{babel}

\usepackage[plainpages=false,colorlinks=true,linkcolor=blue, citecolor=blue, urlcolor=blue]{hyperref}
\usepackage[charter]{mathdesign}
\usepackage[caption=false]{subfig}
\usepackage{amsmath,graphicx,bm}
\usepackage{xcolor}
\usepackage{cleveref}
  
\newcommand\unit{\mathbf{1}}

\newcommand\kv{\mathbf{k}}
\newcommand\rv{\mathbf{r}}
\newcommand\tv{\mathbf{t}}

\newcommand\Mv{\mathbf{M}}
\newcommand\Qv{\mathbf{Q}}
\newcommand\kvt{\mathbf{\tilde k}}

\newcommand\xv{\mathbf{x}}
\newcommand\yv{\mathbf{y}}
\newcommand\Gv{\mathbf{G}}
\newcommand\Ov{\mathbf{O}}

\newcommand\Tv{\mathbf{T}}

\newcommand\Tr{\,\mathrm{Tr}\,}
\newcommand\tr{\,\mathrm{tr}\,}
\newcommand\Sigmav{\bm{\Sigma}}

\renewcommand\t{\theta}
\newcommand\up{\uparrow}
\newcommand\dn{\downarrow}

\newcommand\Gammav{\bm{\Gamma}}
\newcommand\thetav{\bm{\theta}}
\newcommand\epsilonv{\bm{\epsilon}}
\newcommand\eps{\varepsilon}
\newcommand\om{\omega}
\newcommand\dg{\dagger}
\newcommand\dgf{{\phantom{\dagger}}}

\newcommand\etal{\textit{et al.}}

\begin{document}
\title{Coexistence of Superconductivity and Antiferromagnetism in the Hubbard model for cuprates}

\author{A. Foley}
\email[Corresponding author: ]{Alexandre.Foley@usherbrooke.ca}
\affiliation{D\'epartement de physique and Institut quantique, Universit\'e de Sherbrooke, Sherbrooke, Qu\'ebec, Canada J1K 2R1}
\author{S. Verret}\affiliation{D\'epartement de physique and Institut quantique, Universit\'e de Sherbrooke, Sherbrooke, Qu\'ebec, Canada J1K 2R1}
\author{A.-M. S. Tremblay}\affiliation{D\'epartement de physique and Institut quantique, Universit\'e de Sherbrooke, Sherbrooke, Qu\'ebec, Canada J1K 2R1} 
\affiliation{Canadian Institute for Advanced Research, Toronto, Ontario, Canada, M5G 1Z8}
\author{D. S\'en\'echal}\affiliation{D\'epartement de physique and Institut quantique, Universit\'e de Sherbrooke, Sherbrooke, Qu\'ebec, Canada J1K 2R1} 
\date{\today}

\begin{abstract}
Antiferromagnetism and $d$-wave superconductivity are the most important competing ground-state phases of cuprate superconductors.
Using cellular dynamical mean-field theory (CDMFT) for the Hubbard model, we revisit the question of the coexistence and competition of these phases in the one-band Hubbard model with realistic band parameters and interaction strengths.
With an exact diagonalization solver, we improve on previous works with a more complete bath parametrization which is carefully chosen to grant the maximal possible freedom to the hybridization function for a given number of bath orbitals.
Compared with previous incomplete parametrizations, this general bath parametrization shows that the range of microscopic coexistence of superconductivity and antiferromagnetism is reduced for band parameters for $\mathrm{Nd}_{2-x} \mathrm{Ce}_{x} \mathrm{CuO}_{4}$, and confined to electron-doping with parameters relevant for $\mathrm{YBa}_{2} \mathrm{Cu}_{3} \mathrm{O}_{7-\mathrm{X}}$. 
\end{abstract}

\maketitle

\section{Introduction}

The proximity of antiferromagnetism (AF) with superconductivity (SC) is common in unconventional superconductors: Bechgaard salts, heavy-fermion superconductors, high-temperature superconductors (cuprates), iron pnictides and selenides, can all go from antiferromagnetic to superconducting upon varying some parameter (doping, pressure, etc.).
Microscopic, i.e., spatially homogeneous, coexistence of superconductivity with antiferromagnetism is a definite possibility in iron pnictides~\cite{Pratt2009} and selenides~\cite{Liu2011}, in the heavy-fermion compound $\rm CeRhIn_5$~\cite{Knebel2006, Kawasaki2003} and in electron-doped
cuprate superconductors~\cite{Armitage2010}. In this respect, hole-doped cuprates are quite different: the only magnetic phase with which superconductivity coexists is an incommensurate spin-density wave~\cite{yamada_doping_1998,wakimoto_direct_2000,wakimoto_observation_1999,fujita_static_2002}.
A clear difficulty is to distinguish microscopic coexistence (a pure phase) from macroscopic coexistence resulting from inhomogeneities in the sample or from thermodynamic separation of competing phases.
 
The antiferromagnetic phase breaks rotation symmetry ($SO(3)$) and can be characterized by an order parameter $\Mv$, the staggered magnetization. 
Superconductivity, on the other hand, breaks the $U(1)$ symmetry associated with electron number conservation and the associated order parameter is the pairing amplitude $\Delta$.
A signature of the microscopic coexistence of these two phases would be the presence of the so-called $\pi$-triplet order parameter\cite{Demler2004,Almeida2017,Himeda1999}, which is necessarily nonzero if both $\mathbf{M}$ and the $d$-wave order parameter $\Delta$ are nonzero.
Note that the $\pi$-triplet is a kind of pair-density wave~\cite{verret_subgap_2017, lee_amperean_2014, norman_quantum_2018}. 
However, it is different from the pair-density wave observed experimentally in scanning tunnelling microscopy~\cite{hamidian_detection_2016}.
A unified description of the two broken symmetries can be formulated in the language of $SO(5)$ symmetry~\cite{Demler2004}.
A phenomenological Landau-Ginzburg theory of the interplay and coexistence of the two phases can also be formulated without reference to the $SO(5)$ description~\cite{Almeida2017}.

The issue of a possible AF-SC coexistence in high-$T_c$ superconductors has been addressed theoretically using the one-band Hubbard model and its strong-coupling limit, the $t$-$J$ model.
Inui \etal\ found microscopic AF-SC coexistence in a slave-boson (mean field) treatment of the Hubbard model~\cite{Inui1988}.
Himeda \etal\ found it in a variational Monte Carlo study of the $t$-$J$ model~\cite{Himeda1999}. 
The presence of the $\pi$-triplet order parameter was studied in the mean-field approximation by Kyung~\cite{kyung_mean-field_2000}, also in the $t$-$J$ model.
Beyond the mean-field approximation, microscopic AF-SC coexistence was predicted to occur within the Hubbard model with the Variational Cluster Approximation (VCA)~\cite{senechal_competition_2005} and Cluster Dynamical Mean Field Theory (CDMFT)~\cite{capone_competition_2006, kancharla_anomalous_2008}.
In Ref.~\cite{capone_competition_2006}, microscopic coexistence for the nearest-neighbor hopping model was found only for small interaction strength ($U\leq 8t$). 
Functional Renormalization Group (FRG) methods, although more relevant to weak and moderate coupling, also predict the occurrence of such a microscopic coexistence phase~\cite{reiss_renormalized_2007, yamase_coexistence_2016}.

The lack of microscopic coexistence of superconductivity with commensurate antiferromagnetism in hole doped cuprates casts some doubt on the prediction of quantum cluster methods or on the relevance of the one-band Hubbard model to these materials.
In this paper, we show that a more careful application of CDMFT to the one-band Hubbard model makes this AF-SC microscopic coexistence disappear in models relevant to hole-doped cuprates, while reducing its range in a model of electron-doped cuprates.
We use a CDMFT impurity solver based on exact diagonalization at zero temperature, like in Refs~\cite{capone_competition_2006, kancharla_anomalous_2008}, and compare the simple parametrization that they used with the most general parametrization of the bath orbitals, as used in Ref.~\cite{koch_sum_2008,liebsch_finite-temperature_2009, liebsch_temperature_2012}.
Quantum Monte Carlo (QMC) solvers, especially state of the art continuous-time (CT-QMC) solvers~\cite{gull_continuous-time_2011} are free of bath parametrization ambiguities. 
Up to now, CT-QMC solvers have been used to study only the superconducting~\cite{jarrell_phase_2001,maier_kinetic_2004,maier_systematic_2005,Sordi_strong_2012,lin_two-particle_2012,gull_superconductivity_2013,gull_energetics_2012,sordi_c_2013,gull_superconducting_2013,staar_taking_2013,semon_ergodicity_2014,gull_pairing_2014,gull_quasiparticle_2015,fratino_organizing_2016,reymbaut_antagonistic_2016} and the antiferromagnetic phases~\cite{jarrell_phase_2001,fratino_signatures_2017,fratino_effects_2017} separately. 
In principle, the question of coexistence can be addressed with these approaches, but this has yet to be done. 

A QMC cluster size scaling study~\cite{sakai_cluster-size_2012} has demonstrated that $2\times2$ plaquettes give reasonably well converged results and exponential convergence of local observables with cluster size has been observed~\cite{biroli_cluster_2002,aryanpour_comment_2005,biroli_reply_2005}.
Because of this and since a $2\times2$ cluster is already very close to the limit of what is feasible with ED-CDMFT\footnote{The bath structure being simpler in one dimension, computations with more than four correlated orbitals are possible with an ED solver.~\cite{koch_sum_2008}} given the computational resources available to us, we only consider that one size of cluster.   

This paper is organized as follows: 
In Section II we present the model and explain the method used (ED-CDMFT), with a particular attention towards the bath parametrization. 
In Section III we present and discuss our results, before concluding.

\section{Model and method}


Although high-temperature superconductors are charge-transfer insulators~\cite{zaanen_band_1985}, they are often modelled by the one-band Hubbard model on a square lattice:
\begin{equation}\label{eq:Hubbard}
H = -\sum_{\rv,\rv',\sigma} t_{\rv,\rv'} c^\dagger_{\rv,\sigma} c_{\rv',\sigma}
+ U \sum_\rv n_{\rv\up} n_{\rv\dn} - \mu \sum_{\rv,\sigma} n_{\rv\sigma}.
\end{equation}
The hopping amplitudes $t_{\rv,\rv'}$ depend on the particular compound and are restricted to nearest-neighbor ($t$), second-neighbor ($t'$) and third neighbour ($t''$).
We use two sets of parameters: one for $\mathrm{YBa}_{2} \mathrm{Cu}_{3} \mathrm{O}_{7-\mathrm{X}}$ ( YBCO: $t'/t = -0.3$, $t''/t = 0.2$)~\cite{andersen_lda_1995}, a hole-doped compound, and one for $\mathrm{Nd}_{2-x} \mathrm{Ce}_{x} \mathrm{CuO}_{4}$ ( NCCO: $t'/t=-0.17$, $t''/t=0.03$)~\cite{kyung_pseudogap_2004}, an electron-doped compound.
The first neighbour hopping, $t$, defines the energy scale and is set to unity ($t=1$).
The NCCO hoppings can also be considered representative of a class of hole-doped cuprates to which $\mathrm{La}_{2-\mathrm{x}} \mathrm{Sr}_{\mathrm{x}} \mathrm{Cu} \mathrm{O}_{2}$ belongs (see Fig.~5 of Ref.\onlinecite{markiewicz_entropic_2017}).

\subsection{Cluster Dynamical Mean Field Theory}

In CDMFT, for the purpose of computing the electron self-energy $\Sigmav$, the above model is replaced by a cluster model (in this paper, a four-site plaquette) immersed in an effective medium. 
With an exact diagonalization solver, this medium is represented by a finite set of uncorrelated bath orbitals hybridized with the cluster sites.
This discretization of the medium is an additional approximation that must be made to accommodate an ED solver, and as such can lead to additional finite size effects~\cite{koch_sum_2008}.
These bath orbitals, together with the cluster, are described by an Anderson impurity model (AIM):
\begin{equation}\label{eq:impurity}
H_{\rm imp} = H_{\rm clus} + \sum_{\alpha,\xi}\left(\theta_{\alpha\mu}c^\dagger_\alpha a_\mu + \mathrm{H.c.}\right) + \sum_{\mu,\nu} \epsilon_{\mu\nu} a_{\mu}^\dagger a_{\nu}
\end{equation}
where $H_{\rm clus}$ is the restriction of the Hubbard Hamiltonian~\eqref{eq:Hubbard} to the cluster. 
$a_\nu$ annihilates an electron on the bath orbital labelled $\nu$ ($\nu$ stands for both orbital and spin indices and so does $\alpha$ for cluster orbitals).
The matrix $\theta_{\alpha\mu}$ defines the hybridization between bath and cluster, and $\epsilon_{\mu\nu}$ defines the dynamics of the bath.
The bath parameters $\theta_{\alpha\mu}$ and $\epsilon_{\mu\nu}$ are determined by an iterative procedure, as explained below.

The one-electron Green function $\Gv'$ takes the following form as a function of complex frequency $\om$:
\begin{equation}\label{eq:sigma}
\Gv'^{-1}(\om) = \om - \tv - \Gammav(\om) - \Sigmav(\om)
\end{equation}
where the hybridization matrix $\Gammav(\om)$ is defined as
\begin{equation}\label{eq:hybridization}
\Gammav(\om) = \thetav {(\om - \epsilonv)}^{-1}\thetav^\dagger
\end{equation}
in terms of the matrices $\theta_{\alpha\mu}$ and $\epsilon_{\alpha\beta}$.
In practice, the cluster Green function is computed from an exact diagonalization technique and the self-energy is extracted from Eq.~\eqref{eq:sigma}.

The Green function $\Gv(\kvt,\om)$ for the original lattice Hubbard model is then computed from the cluster’s self-energy as
\begin{equation}\label{eq:latticeGF}
\Gv^{-1}(\kvt,\om) = \Gv_0^{-1}(\kvt,\om) - \Sigmav(\om)~.
\end{equation}
Here $\kvt$ denotes the wave vectors belonging to the Brillouin zone associated with the superlattice of plaquettes, and $\Gv_0$ is the non-interacting lattice Green function.
All Green function-related quantities are $2N_c\times 2N_c$ matrices, $N_c=4$ being the number of sites in the plaquette. 

The bath parameters are ideally determined by the self-consistency condition $\Gv'(\om) = \bar\Gv(\om)$, where
\begin{equation}\label{eq:Gbar}
\bar\Gv(\om) \equiv \int\frac{d^2\tilde k}{{(2\pi)}^2} \Gv(\kvt,\om)
\end{equation} 
and where the integral is carried over the reduced Brillouin zone (the domain of $\kvt$).
In other words, the local Green function $\Gv'(\om)$ should coincide with the Fourier transform $\bar\Gv(\om)$ of the full Green function at the origin of the superlattice. 
This condition should hold at all frequencies, which is impossible in ED-CDMFT because of the finite number of bath parameters $\epsilonv$ and $\thetav$ at our disposal. 
Thus, the self-consistency condition is replaced by a minimization of the so-called \textit{distance function}:
\begin{equation}\label{eq:dist}
d= \sum_{\om_n} W(i\om_n) \Tr  \Big\vert\Gv'^{-1}(i\om_n) -\bar\Gv^{-1}(i\om_n) \Big\vert^2\mathrm{,}
\end{equation} 
with respect to the bath parameters for a given value of $\Sigmav$. The weight function $W(x)$ is meant to give more importance to small frequencies and a fictitious temperature is used to define the grid of Matsubara frequencies over which the above merit function is evaluated. 
Details can be found, for instance, in Refs~\cite{caffarel_exact_1994, kancharla_anomalous_2008,senechal_bath_2010,avella_cluster_2012}.
In this work, we use a fictitious temperature $\beta=50/t$ and set $W(i\om_n)$ to 1 if $\om_n<\om_c$, with $\om_c=2t$;
$W(i\om_n)$ is set to zero for higher Matsubara frequencies.
 
To summarize, 
the ED-CDMFT procedure runs as follows:
\begin{enumerate}
\item An initial, trial set of bath parameters is chosen.
\item The ED solver is applied to Hamiltonian~\eqref{eq:impurity} and provides a numerical representation of $\Gv'$ that allows for a quick computation of $\Gv'(\om)$, and hence of $\Sigmav(\om)$, at any complex frequency.
\item The Fourier transform $\bar\Gv(\om)$ is computed for the set of Matsubara frequencies appearing in~\eqref{eq:dist}.
\item A new set of bath parameters $\{\theta_{\alpha\mu}, \eps_{\mu\nu}\}$ is obtained by minimizing the distance function~\eqref{eq:dist}.
\item We go back to step 2 with the new set of bath parameters until convergence is achieved.
\end{enumerate}

Note that superconductivity is allowed in the procedure through the use of the Nambu formalism, by which a particle-hole transformation is performed on the spin-down orbitals (e.g. $c_{\alpha,\downarrow} \rightarrow c^\dag_{\alpha,\downarrow}$). This does not require a doubling of the degrees of freedom in the Green function if no spin-flip terms are present in the model, which is the case here.
With 8 bath orbitals, we introduce two Nambu spinors, one for the cluster and one for the bath:
\begin{align}
C &= (c_{1\up}\ldots c_{4\up},c^\dg_{1\dn}\ldots c^\dg_{4\dn}) \\
A &= (a_{1\up}\ldots a_{8\up},a^\dg_{1\dn}\ldots a^\dg_{8\dn})~.
\end{align}
The noninteracting part of Hamiltonian~\eqref{eq:impurity} then takes the general form
\begin{equation}\label{eq:Himp0}
H_{\rm imp, 0} = (C^\dg, A^\dg)\begin{pmatrix}
\Tv & \thetav \\ \thetav^\dg & \epsilonv 
\end{pmatrix} \begin{pmatrix}
C\\ A
\end{pmatrix}
\end{equation}
where the matrices $\thetav$ and $\epsilonv$ can now contain anomalous terms.

Two different bath parametrizations are used in this paper, as described in the next section.
They parametrize the same number of bath orbitals, but they differ in the number of free parameters that are set by the CDMFT procedure.
The first bath parametrization is essentially the same as the one used in Ref.~\cite{kancharla_anomalous_2008}; the second, inspired by Ref.~\cite{liebsch_finite-temperature_2009,koch_sum_2008}, contains many more independent parameters.
One may naturally expect that increasing the number of bath parameters brings the system closer to perfect self-consistency. 
Using the second parametrization gives us the best possible self-consistency for a given number of orbitals; this is where this paper improves upon previous studies.
On the other hand, the CDMFT procedure itself becomes more time consuming, in particular optimizing the merit function Eq.~\eqref{eq:dist}.

\subsection{Simple bath parametrization}\label{sec:simple}

\begin{figure}
	\subfloat[$ $]
	{\label{fig:bath_simple:a}
		\includegraphics[width=0.67\hsize]{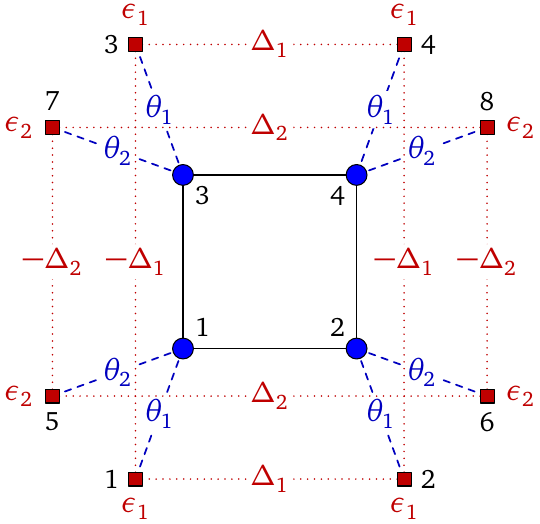}
	}
	\vspace{0cm}
	\subfloat[$ $]
	{\label{fig:bath_simple:b}%
		\includegraphics[width=0.76\hsize]{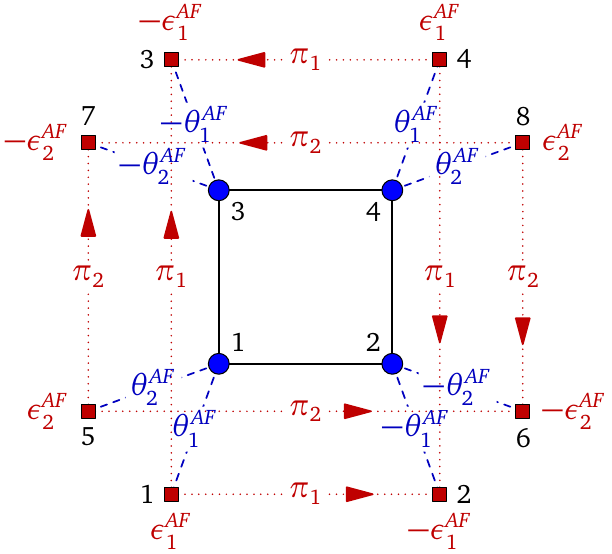}
	}
\caption{Schematic description of the simple bath parametrization. 
The blue circles represent the cluster sites and red squares the bath orbitals.
Intra bath terms are shown in red and hybridization terms in blue.
Cluster and bath orbitals are numbered separately.
(a): terms that are used when not probing antiferromagnetism, with $C_{2v}$ symmetry. (b): additional terms needed when probing antiferromagnetism and breaking $C_{2v}$ down to $C_2$. See text for details.}
\label{fig:bath_simple}
\end{figure}

A simple and intuitive way to configure the bath orbitals for a 4-site plaquette is illustrated on Fig.~\ref{fig:bath_simple}.
Time and memory constraints limit the number of bath orbitals to 8, for a total of 12 orbitals in the AIM Eq.~\eqref{eq:impurity}.
The 8 bath orbitals are separated into two sets, numbered 1 and 2, each with four orbitals.
Each set can be thought of as a ``ghost'' of the plaquette, with a site energy $\eps_i$ ($i=1,2$), hybridized with the bath through a hopping amplitude $\theta_i$.
The presence of superconductivity in the effective medium is characterized by a singlet pairing amplitude $\Delta_i$ with opposite signs along the $x$ and $y$ directions, in accordance with $d$-wave symmetry.
This makes a total of 6 independent parameters, as summarized on Fig.~\ref{fig:bath_simple:a}.
If only superconductivity is probed, the AIM has a $C_{2v}$ symmetry (horizontal and vertical reflexions) and only those 6 bath parameters are used.

If antiferromagnetism is considered as well, then the symmetry is reduced to $C_2$ (a $\pi$ rotation). 
Six additional parameters are introduced, as illustrated on Fig.~\ref{fig:bath_simple:b}:
Spin antisymmetric bath energies $\eps_i^{AF}$ and hybridizations $\theta_i^{AF}$ that alternate in sign between sites,
and triplet-pairing amplitudes $\pi_i$ whose signs are defined on the figure (via arrows).
This makes a total of 12 independent bath parameters. 
 
Let us specify the matrices introduced in Eq.~\eqref{eq:Himp0}.
We order the bath orbitals so the two sets of 4 spin-up orbitals are consecutive and followed by the two sets of spin down orbitals, in the same order.
Overall, the $16\times16$ matrix $\epsilonv$ associated with this bath model has the following structure, in terms of $4\times4$ blocks:
\begin{equation}
\epsilonv = \begin{pmatrix}
E_1+E_1^{AF} & 0 & D_1 & 0 \\
0 & E_2+E_2^{AF} & 0 & D_2 \\
D_1 & 0 & -E_1+E_1^{AF} & 0 \\
0 & D_2 & 0 & -E_2+E_1^{AF} \\
\end{pmatrix}
\end{equation}
where $E_i = \eps_i\unit_{4\times4}$, $E_i^{AF} = \eps_i^{AF}\mathrm{diag}(1, -1, -1, 1)$
and
\begin{equation}
D_i = \begin{pmatrix}
0	& -\Delta_i+\pi_i	& \Delta_i-\pi_i	& 0	 \\
-\Delta_i-\pi_i	& 0	& 0	& \Delta_i+\pi_i	 \\
\Delta_i+\pi_i	& 0	& 0	& -\Delta_i-\pi_i	 \\
0	& \Delta_i-\pi_i	& -\Delta_i+\pi_i	& 0	
\end{pmatrix}~.
\end{equation}
The minus signs in the bottom half of the $\boldsymbol{\epsilon}$ matrix comes from the Nambu representation of spin down orbitals.
On the other hand, the $8\times16$ matrix $\thetav$ has the following structure:
\begin{equation}
\thetav = \begin{pmatrix}
\Theta_1+\Theta_1^{AF} & \Theta_2+\Theta_2^{AF} & 0 & 0 \\
0 & 0 & -\Theta_1+\Theta_1^{AF} & -\Theta_2+\Theta_2^{AF} 
\end{pmatrix}
\end{equation}
where $\Theta_i = \theta_i\unit_{4\times4}$, $\Theta_i^{AF} = \theta_i^{AF}\mathrm{diag}(1, -1, -1, 1)$.

\subsection{General bath parametrization}\label{sec:general}

\begin{figure}
 \subfloat[$C_{2v}$]{
 \label{fig:Param:a}%
 \centering
 \hspace{-0.02\hsize}
  \includegraphics[scale=0.95]{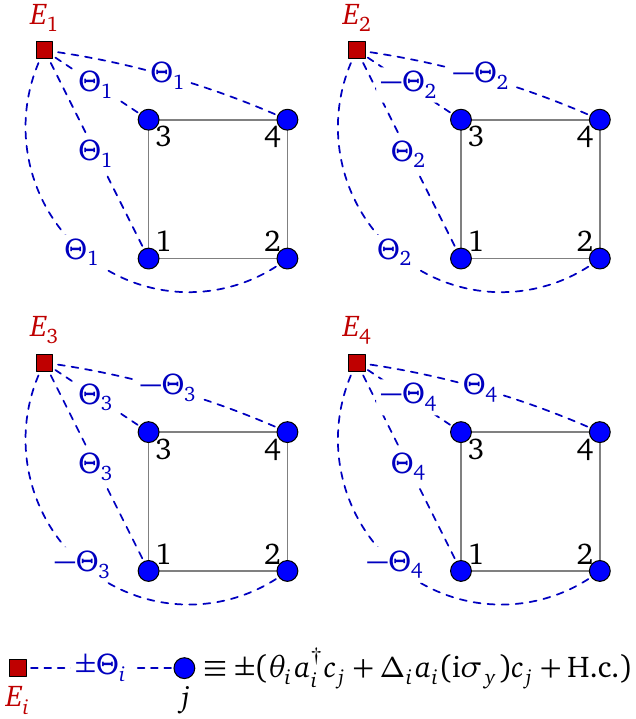}
	}

 \subfloat[$C_{2}$]{\label{fig:Param:b}%
 \centering
  \includegraphics[scale=0.95]{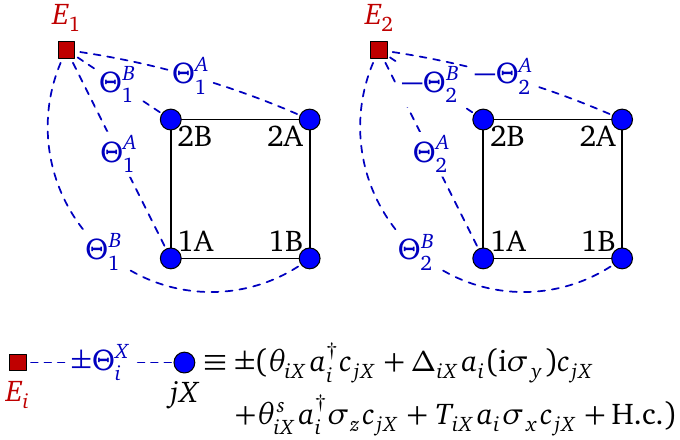}
	}

  \caption{
Top (a): Schematic description of the general parametrization of the bath when probing superconductivity with $C_{2v}$ symmetry.
Blue circles represent the cluster sites and red squares the bath orbitals. The 4 subfigures correspond to the first four bath orbitals, each associated with a different irreducible representation of $C_{2v}$.
The blue dashed lines and accompanying symbols represent a hybridization operator as defined below the figure.
The four diagrams correspond to the first 4 columns (or the last 4 columns) of the expressions~\eqref{theta_block} and~\eqref{delta_block} in the text. 
Bottom (b): the same, for $C_2$ symmetry, used when probing antiferromagnetism and superconductivity.
The operators symbolized by the dashed line are once again defined below the figure.
The superscripts $A$ and $B$ correspond to the two sublattices when AF order is present. In the equations of (a) and (b), spinor notation is used: the Pauli matrices define the spin part of each term.
}
  \label{fig:Param}
\end{figure}

The bath parametrization defined above is not the most general one.
In order to construct the most general bath appropriate to each point group ($C_{2v}$ and $C_2$), one must first realize that any unitary transformation of the bath orbitals is possible, and that this gauge freedom must be fixed somehow. 
We do this by requiring (1) that the matrix $\epsilonv$ be diagonal and (2) that it falls into irreducible representations of the symmetry group of the AIM.\@
 
If only superconductivity is probed, we can impose $C_{2v}$ symmetry (horizontal and vertical reflexions) on the AIM.\@
In addition to $C_{2v}$ symmetry, we also assume spin symmetry, which reduces the number of independent parameters.
$C_{2v}$ has four distinct irreducible representations
to each of which we associate two bath orbitals orbitals, for a total of 8, the same number as in the simple parametrization of the previous section.
We label these orbitals from 1 to 8, in two consecutive series of four, each series corresponding to the four irreducible representations: 1 and 5 in the first representation, 2 and 6 in the second, 3 and 7 in the third, and 4 and 8 in the fourth.
Fig.~\ref{fig:Param:a} illustrates the hybridization of the first four orbitals.
The coefficients $\theta_i$ and $\Delta_i$ are the parameters present in the hybridization between the cluster and the $i$th bath orbital: $\theta_i$ is the amplitude of the simple hoppings and $\Delta_i$ is the amplitude of the singlet pairing operators. 

In this bath parametrization, the matrix $\epsilonv$ is diagonal: $\mathrm{diag}(\eps_i)\oplus\mathrm{diag}(-\eps_i)$, with $i=1,\dots,8$.
Again, the last 8 components correspond to the spin-down orbitals in the Nambu representation (hence the minus sign).

The hybridization matrix $\thetav$, on the other hand, is a dense $8\times16$ matrix:
\begin{equation}
\thetav = \begin{pmatrix}
\Theta & -D\\
D & -\Theta \\
\end{pmatrix}
\end{equation}
where the $4\times8$ blocks $\Theta$ and $D$ are defined as
\begin{align}\label{theta_block}
\Theta &= \begin{pmatrix}
\t_1	& \t_2	& \t_3	& \t_4	& \t_5	& \t_6	& \t_7	& \t_8		\\
\t_1	& \t_2	& -\t_3	& -\t_4	& \t_5	& \t_6	& -\t_7	& -\t_8		\\
\t_1	& -\t_2	& \t_3	& -\t_4	& \t_5	& -\t_6	& \t_7	& -\t_8		\\
\t_1	& -\t_2	& -\t_3	& \t_4	& \t_5	& -\t_6	& -\t_7	& \t_8
\end{pmatrix}
\\
\label{delta_block} 
D &= \begin{pmatrix}
\Delta_1	& \Delta_2	& \Delta_3	& \Delta_4	& \Delta_5	& \Delta_6	& \Delta_7	& \Delta_8		\\
\Delta_1	& \Delta_2	& -\Delta_3	& -\Delta_4	& \Delta_5	& \Delta_6	& -\Delta_7	& -\Delta_8		\\
\Delta_1	& -\Delta_2	& \Delta_3	& -\Delta_4	& \Delta_5	& -\Delta_6	& \Delta_7	& -\Delta_8		\\
\Delta_1	& -\Delta_2	& -\Delta_3	& \Delta_4	& \Delta_5	& -\Delta_6	& -\Delta_7	& \Delta_8
\end{pmatrix}
\end{align}
The number of independent bath parameters is 8 in $\epsilonv$ and $2\times8=16$ in $\thetav$, for a total of 24.

If antiferromagnetism is probed as well, the symmetry reduces to $C_2$, which has only two irreducible representations. 
Since we can afford 8 bath orbitals, we associate 4 bath orbitals to each irreducible representation, with the same pattern as the $C_{2v}$ bath.
Namely, all the odd labelled bath orbitals have the same structure as the first orbital, and all the even bath orbitals have the same structure as the second orbital.
The cluster-bath couplings of the first and second bath orbitals are illustrated in Fig.~\ref{fig:Param:b}. 
The superscripts $A$ and $B$ refer to the two sublattices induced by 
antiferromagnetic order.
The parameters associated with different sublattices may differ.

The energy $E_i$ ($i=1,\ldots,8$) has a component $\eps_i$ even in spin and a component $\eps_i^s$ odd in spin, which makes
$2\times8=16$ parameters.
Thus, the diagonal matrix $\epsilonv$ has the structure
\begin{equation}
(\eps_1+\eps^s_1, \ldots,\eps_8+\eps^s_8,-\eps_1+\eps^s_1,\ldots,-\eps_8+\eps^s_8)
\end{equation}

The operators $\Theta^{A}_i$ in Fig.~\ref{fig:Param:b} contain four parameters each: $\theta^{A}_i$, $\theta^{sA}_i$, $\Delta^{A}_i$,  and $T^{A}_i$, where
$\theta^{A}_i$ is a spin-symmetric hopping operator, $\theta^{sA}_i$ a spin antisymmetric hopping, 
$\Delta^{A}_i$ a singlet pairing and $T^{A}_i$ a triplet pairing; likewise for $\Theta^{B}_i$.
This makes 8 parameters for each bath index $i$, therefore $8\times8=64$ hybridization parameters in total.
The hybridization matrix $\thetav$ has the following form:
\begin{equation}
\thetav = \begin{pmatrix}
\Theta+\Theta^s & -D+T\\
D+T & -\Theta+\Theta^s \\
\end{pmatrix}
\end{equation}
where the $4\times8$ block $\Theta$ is defined as
\begin{equation*}
\Theta = \begin{pmatrix}
\t^A_1	& \t^A_2	& \t^A_3	& \t^A_4	& \t^A_5	& \t^A_6	& \t^A_7	& \t^A_8		\\[2pt]
\t^B_1	& \t^B_2	& \t^B_3	& \t^B_4	& \t^B_5	& \t^B_6	& \t^B_7	& \t^B_8		\\[2pt]
\t^B_1	& -\t^B_2	& \t^B_3	& -\t^B_4	& \t^B_5	& -\t^B_6	& \t^B_7	& -\t^B_8		\\[2pt]
\t^A_1	& -\t^A_2	& \t^A_3	& -\t^A_4	& \t^A_5	& -\t^A_6	& \t^A_7	& -\t^A_8
\end{pmatrix}
\end{equation*}
and likewise for $\Theta^s$ (with components $\t^{sA}_i, \t^{sB}_i$), $D$ (with components $\Delta^A_i$, $\Delta^B_i$) and $T$(with components $T^A_i$, $T^B_i$).
The total number of independent bath parameters in this parametrization is $16+64=80$.
Note that any hybridization function that can be produced with the simple parametrization can also be produced by the general parametrization we described.
By applying an unitary transformation on the simple parametrization, we could obtain a bath Hamiltonian with the same structure as the general parametrization, except for a large number of additional constraints between bath parameters.

\begin{figure*}
\includegraphics[scale=0.8]{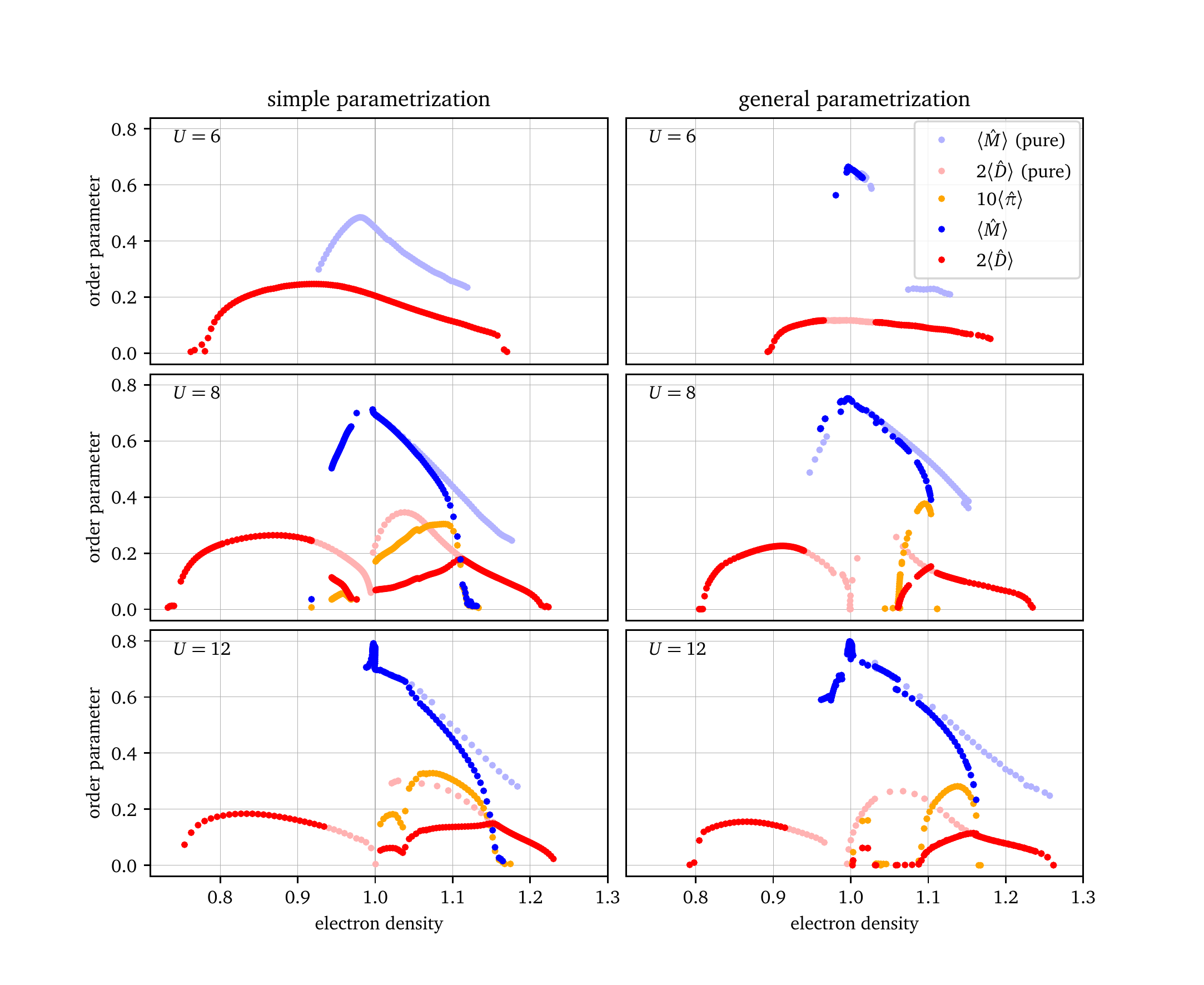}
\caption{Phase diagrams (order parameter as a function of electron density) obtained with the simple (left) and general (right) bath parametrizations, with YBCO-like band parameters ($t'/t = -0.3$, $t''/t=0.2$) and three values of onsite repulsion $U$. Blue symbols represent the AF order parameter $\langle\hat M\rangle$, red symbols the dSC order parameter $\langle\hat D\rangle$ (times 2) and the orange symbols the $\pi$-triplet order (times 10). 
Dark symbols are obtained when allowing microscopic coexistence of the two orders. 
Pale symbols are obtained when probing pure solutions.
Deep in the superconducting regime, the dark symbols can be from pure SC simulations;
allowing for AFM there would have significantly increased computational time with no benefits. 
}
\label{fig:YBCO}
\end{figure*} 

\begin{figure}
\includegraphics[scale=0.8]{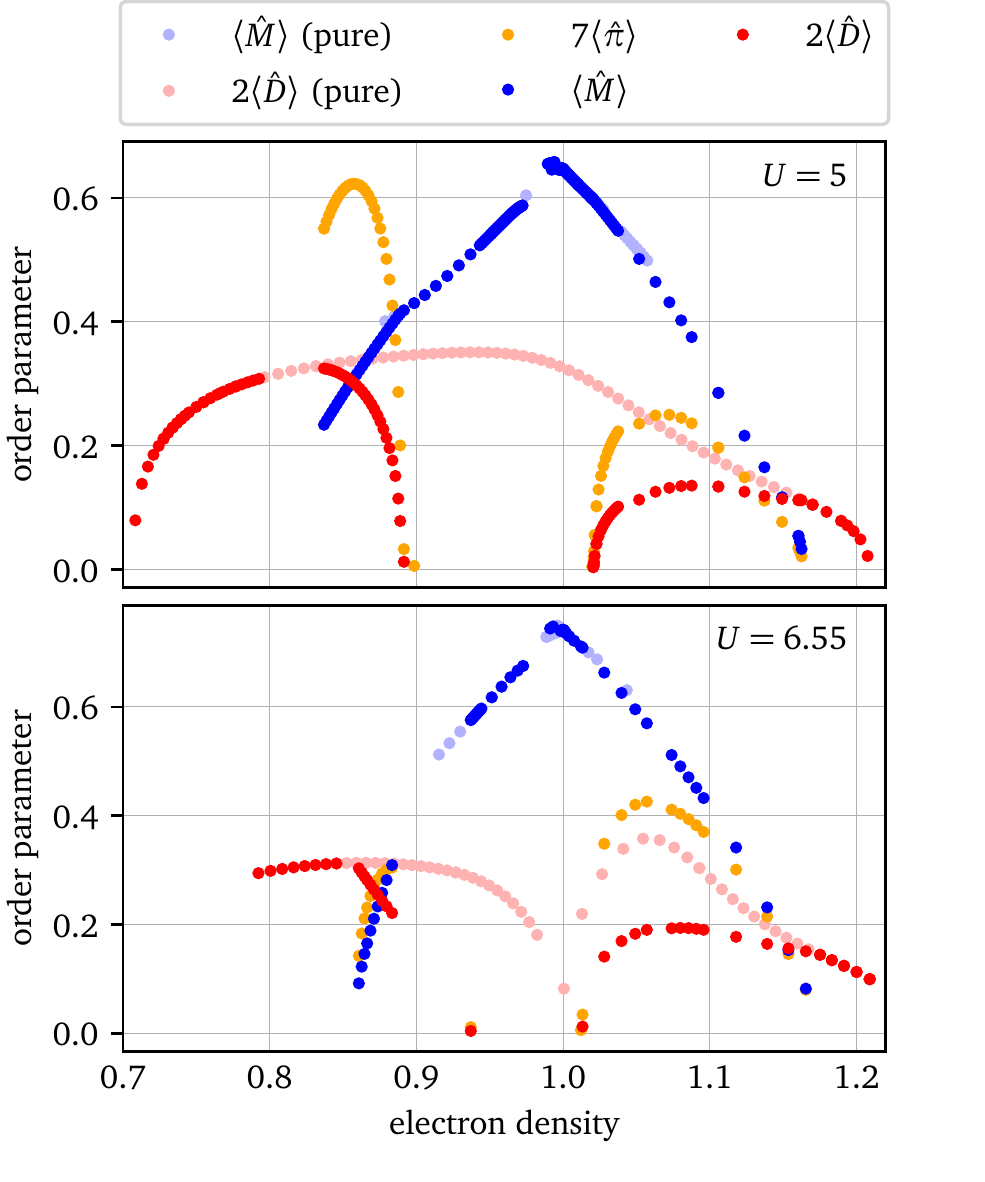}
\caption{Phase diagrams (order parameter as a function of electron density) obtained with the general bath representations, with NCCO-like band parameters ($t'/t = -0.17$, $t''/t=0.03$) and two values of onsite repulsion $U$. Symbols have the same meaning as on Fig.~\ref{fig:YBCO}. Note that the dSC order parameter is multiplied by 2 and the $\pi$-triplet order parameter by 7, for clarity.
}
\label{fig:NCCO}
\end{figure}

\subsection{Order parameters}

Once a CDMFT solution is found, various order parameters may be computed.
The quantities of interest are the averages of the following operators defined on the lattice:
\begin{align}
\hat M &= \sum_\rv e^{i\Qv\cdot\rv} \left[c^\dagger_{\rv\uparrow}c^\dgf_{\rv\uparrow} - c^\dagger_{\rv\downarrow}c^\dgf_{\rv\downarrow}
\right] \qquad \Qv=(\pi,\pi) \\
\hat D &= \sum_\rv \left[ c_{\rv\uparrow}c_{\rv+\xv\downarrow} - c_{\rv\downarrow}c_{\rv+\xv\uparrow} - c_{\rv\uparrow}c_{\rv+\yv\downarrow} + c_{\rv\downarrow}c_{\rv+\yv\uparrow} \right] + \mathrm{H.c} \\
\hat \pi &= \sum_\rv e^{i\Qv\cdot\rv} \left[ c_{\rv\uparrow}c_{\rv+\xv\downarrow} + c_{\rv\downarrow}c_{\rv+\xv\uparrow} - c_{\rv\uparrow}c_{\rv+\yv\downarrow} - c_{\rv\downarrow}c_{\rv+\yv\uparrow} \right] + \mathrm{H.c}~.
\label{eq:pi-triplet}
\end{align}
The first is just the spin-density operator with the antiferromagnetic wavevector $\Qv$. 
The second defines the $d$-wave pairing operator: singlet pairing on nearest-neighbour bonds $\xv$ and $\yv$ with opposite signs.
The third is the $\pi$-triplet operator: triplet pairing on nearest-neighbour bonds with opposite signs along $\xv$ and $\yv$ and a spatial modulation defined by the antiferromagnetic wavevector $\Qv$.

In ED-CDMFT, there are two ways to estimate the average of one-body operators. The first, and also the simplest, consists in computing the expectation value of the restriction of these operators to the cluster in the ground state of the impurity model.
The averages obtained in this way will be called \textit{cluster averages}.

The second method involves the lattice model Green function~\eqref{eq:latticeGF}.
Specifically, the average of any one-body operator of the form $\hat O = O_{\alpha\beta} c^\dagger_\alpha c_\beta$
can be computed from the Green function as 
\begin{equation}
\langle \hat O\rangle = \oint\frac{d\omega}{2\pi} \tr[\Ov\Gv(\omega)]~.
\end{equation}
The frequency integral is taken along a contour that surrounds the negative real axis.
In practice, this is done in the mixed basis of superlattice wavevectors $\tilde\kv$ and cluster orbitals indices, knowing that both $\Ov$ and $\Gv$ are diagonal in $\kvt$:
\begin{equation}\label{eq:average1}
\langle \hat O\rangle = \oint\frac{d\omega}{2\pi}\int\frac{d^2\tilde k}{{(2\pi)}^2} \tr[\Ov(\kvt)\Gv(\kvt, \omega)]
\end{equation}
The averages obtained in this way will be called \textit{lattice averages}.

An operator $\hat O$ that is defined on sites only, not on bonds, like $\hat M$ above, or the electron density $\hat n$, will be called a \textit{local operator}.
For such an operator $\Ov(\kvt)$ does not depend on $\kvt$ and the above formula simplifies to
\begin{equation}\label{eq:average2}
\langle \hat O\rangle = \oint\frac{d\omega}{2\pi} \tr[\Ov\bar\Gv(\omega)]
\end{equation}
where the local Green function $\bar\Gv$ is defined in Eq.~\eqref{eq:Gbar}.
For a local operator, the cluster average, instead of being computed from the impurity ground state, could alternatively be computed from Eq.~\eqref{eq:average2}, but with the impurity Green function $\Gv'$ substituted for $\bar\Gv$, which yields the cluster average mentioned before.
	
Cluster and lattice averages are not equal for two reasons. First, operators that live on bonds, like $\hat D$ and $\hat\pi$, differ from their restrictions to the cluster: inter-cluster bonds are ignored. Lattice averages take these inter-cluster bonds into account, cluster averages do not.
Second, in ED-CDMFT the self-consistency is only approximate; therefore the local Green function $\bar\Gv$ is not exactly the same as the impurity Green function $\Gv'$. Thus even for a local operator, for which Eq.~\eqref{eq:average2} applies, the lattice average will be only approximately the same as the cluster average.

\section{Results}

In this section we present zero-temperature phase diagrams: the order parameters as a function of density, for two sets of band parameters and five values of $U$.
We display cluster averages for local operators (electron density $\hat n$ and staggered magnetization $\hat M$) and lattice averages for bond operators ($\hat D$ and $\hat\pi$).
See the supplementary material for other combinations and for comments on the differences between them~\cite{sup_mat}. 
The lowest value of $U$ for each set (YBCO- and NCCO-like) lies below the Mott transition; hence superconductivity, when probed alone, extends all the way to half-filling.
The other values of $U$ are above the Mott transition and hence superconductivity vanishes exactly at half-filling in the underdoped regime.  
In larger clusters, both in dynamical cluster approximation with a Quantum Monte Carlo solver\cite{gull_superconductivity_2013} and in exact diagonalization with the variational cluster approximation~\cite{guillot_competition_2007}, it is found that $d$-wave superconductivity (dSC) starts away from half-filling. 

\subsection{YBCO-like parameters}\label{ss:YBCOLP}

Fig.~\ref{fig:YBCO} shows the AF and dSC order parameters obtained in CDMFT for a one-band dispersion appropriate to YBCO and three values of $U$ ($6$, $8$ and $12$), as a function of electron density $n$ (half-filling corresponds to $n=1$). 
The left panels show the solutions obtained using the simple parametrization of Sect.~\ref{sec:simple} and the right panels the solutions obtained using the general parametrization of Sect.~\ref{sec:general}. 
Even though YBCO is a hole-doped compound, we have obtained solutions on both hole- and electron-doped sides of the phase diagram.

Solutions were obtained either by allowing both orders to emerge simultaneously in microscopic coexistence, or by allowing only one order at a time, suppressing the other.
For instance, in the simple parametrization, antiferromagnetism is suppressed by forcing the bath parameters illustrated in the lower panel of Fig.~\ref{fig:bath_simple} to zero.
Similarly, suppressing superconductivity is done by setting to zero all bath pairing operators ($\Delta_{1,2}$ and $\pi_{1,2}$). 
Even when both orders are allowed, microscopic coexistence does not necessarily happen: one of the two orders (AF or dSC) may dominate, in which case the variational parameters of the other order (dSC or AF) reach zero by the end of the self-consistency loop.
To construct our phase diagrams, we let the self-consistency procedure choose which phase is the right one, rather than comparing ground-state energy. We proceed this way because thermodynamic potentials are unreliable with a small finite bath.

In Figs.~\ref{fig:YBCO} and \ref{fig:NCCO}, dark symbols indicate the order parameter, blue for AF and red for dSC, in the microscopically coexistent or dominant solution.
Pale symbols indicate subdominant solutions, i.e., solutions obtained by suppressing one order in the coexisting regime.
Orange circles denote the average of the $\pi$-triplet operator of Eq.~\eqref{eq:pi-triplet}, which is nonzero in regions of microscopic AF-dSC coexistence.

One notices the following features:
\begin{enumerate}
\item Microscopic coexistence does not occur at $U=6$, in both simple and general bath parametrizations. It occurs only at $U=8$ and, in a wider range of doping, at $U=12$.
\item In the general parametrization, microscopic coexistence only occurs on the electron-doped side, whereas it also occurs on the hole-doped (i.e. physical) side in the simple parametrization.
\item The pure antiferromagnetic solutions show many discontinuities, especially on the hole-doped side and at stronger coupling, and especially in the more general parametrization.
\item Where microscopic coexistence occurs with the general parametrization, the transitions from a pure phase to microscopic coexistence are of second order.
\item A small superconducting region can be seen around $n=1.01$ at $U=12$ with the general parametrization. This is a finite size effect due to a change in the number of particles in the AIM. On either side of this small dome, the AIM has a well defined number of particles. The small superconducting order parameter breaks particle number conservation and allows the change in the number of particles to happen smoothly over a finite range of doping, instead of abruptly as it would were particle number conservation enforced.

\item On the hole-doped side, the transition from superconductivity to antiferromagnetism is of first order, and the two solutions are separated in density. In principle, this leads to a macroscopic phase coexistence.
\end{enumerate}

\subsection{NCCO-like parameters}

Fig.~\ref{fig:NCCO} shows the same quantities for band parameters adequate for NCCO and two values of $U$.
One can make the following observations:

\begin{enumerate}
\item Microscopic coexistence occurs on both sides of the phase diagram, for both values of $U$ considered.
\item On the hole-doped side, the transition from microscopic coexistence to pure SC (upon decreasing $n$) is discontinuous for $U=5$; the two solutions (pure SC and microscopic coexistence) are separated in density, which leads in principle to macroscopic phase coexistence.
Macroscopic coexistence between a pure SC phase and a microscopic SC-AF coexistence phase has been seen before in VCA for different band parameters~\cite{aichhorn_antiferromagnetic_2006}.
\item On the hole-doped side, the transition from microscopic coexistence to pure AFM (upon increasing $n$) is discontinuous at $U=6.55$ and continuous at $U=5$.
\item On the electron doped (i.e. physical) side, the transition to microscopic coexistence is continuous for both values of $U$.
\end{enumerate} 

\section{Discussion}

On general grounds, the phase diagrams of Fig.~\ref{fig:YBCO} obtained with the general parametrization should be more representative of the result obtained with an infinite bath; in other words, closer to an accurate solution of the Hubbard model.
Since we find that the results for this general parametrization are closer to the experimental phase diagram of cuprates, the appropriateness of the one-band Hubbard model for cuprate superconductors is reinforced.

On physical grounds, we can argue that the phase diagrams obtained with the general parametrization are more accurate.
Consider $U$ slightly below the critical value for the Mott transition, in our case at $U=6$: the simple parametrization leads to a superconducting ground state at half-filling, whereas the general parametrization favor antiferromagnetism. 
The latter result is more sensible considering antiferromagnetism at half-filling can be obtained at the Hartree-Fock level while $d$-wave superconductivity is a dynamical effect.
Another example is the reentrant behavior of superconductivity upon underdoping obtained with the simple parametrization at $U=12$, between half-filling and $4\%$ electron doping: it is inexplicable on physical grounds.
This reentrant behavior is suppressed when using the general parametrization: there is a slight reduction in amplitude and it appears at lower electron doping, before $3\%$.
It also becomes well separated from the superconducting dome. 
This distance from the dome allows for a meaningful analysis of the particle number of the AIM on each side of this reentrant feature and leads us to believe that it is a finite-size effect, as pointed out in section~\ref{ss:YBCOLP}.

One can observe in Fig.~\ref{fig:YBCO} that, as $U$ increases, the difference between the phase diagrams of the two parametrizations becomes more subtle, especially on the hole doped side. 
At $U=12$, the results are qualitatively equivalent on the hole-doped side. 
We can only speculate on why this happens.
As the general parametrization can reproduce any hybridization function the simple one can generate, it is possible the converged hybridization functions produced by the two parametrizations are more similar at higher $U$ values. This would mean that the additional constraints of the simple parametrization become less of an issue as the states become more localized.
In other words, it could mean that some of the constraints of the simple parametrization are physically meaningful at strong coupling. 
 
Let us note that even our ``simple parametrization'' has more variational parameters than what was used in previous ED-CDMFT studies~\cite{kancharla_anomalous_2008, capone_competition_2006}.
In these previous studies, the $\pi$-triplet had no associated variational parameter in the AIM Hamiltonian, even if its expectation value does not vanish in the microscopic coexistence phase.
These additional degrees of freedom allow for slightly stronger $d$-wave superconductivity, although the overall qualitative shape of the phase diagram is unchanged. One should also note that Ref.~\onlinecite{capone_competition_2006} considers a particle-hole symmetric lattice, leading to notably different phase diagrams. 

At intermediate coupling, the regime relevant for cuprates, details of the band structure, the value of $t'$ in particular, are just as important as the interaction strength to determine the phase diagram, as noted in Ref.~\onlinecite{kancharla_anomalous_2008}.
Indeed, Figure~\ref{fig:YBCO}, with band parameters relevant to YBCO, shows that electron-hole symmetry is strongly violated. 
In the general parametrization, microscopic coexistence between $d$-wave superconductivity and antiferromagnetism is confined to the electron-doped region. 
On the hole-doped side, the transition between antiferromagnetism and $d$-wave superconductivity is of first order for $U=8$ and $U=12$, for both parametrizations. 
Note that the wider range of doping for antiferromagnetism on the electron-doped side simply reflects the better nesting at the antiferromagnetic wave vector~\cite{senechal_competition_2005}. 

It is quite remarkable that the electron-doped side realizes a proposal by Sachdev~\cite{sachdev_quantum_2010} that the presence of $d$-wave superconductivity leads to a large displacement of the doping at which antiferromagnetism ends. 
This can be seen by comparing the end of the pale blue dots for $U=8$ or $12$ with the end of the dark blue dots in Fig.~\ref{fig:YBCO}. However, Sachdev’s conjecture concerned the hole-doped compound where we do not observe this effect.

A continuous time QMC computation~\cite{fratino_effects_2017} of the antiferromagnetic phase with $t'=-0.1$ shows that it extends to $~15\%$ hole doping, like what we can see in Fig.~\ref{fig:NCCO}. This suggests that, with these band parameters and coupling, the presence of superconductivity has very little effect on the antiferromagnetic phase.
An FRG study of microscopic coexistence~\cite{reiss_renormalized_2007} at weaker coupling but similar band parameters finds a hole doped phase diagram strikingly similar to Fig.~\ref{fig:NCCO} at $U=5$. This shows that those results are robust when long wavelengths and incommensurate orders are suppressed.

Our small clusters cannot sustain waves with periods longer than two unit cells, like the charge order seen in experiments~\cite{comin_resonant_2016} or the incommensurate spin-waves seen in both experiments~\cite{haug_neutron_2010,wakimoto_direct_2000,yamada_doping_1998,wakimoto_observation_1999,fujita_static_2002} and infinite lattice weak-coupling calculations~\cite{yamase_coexistence_2016,eberlein_fermi_2016,vilk_non-perturbative_1997,schulz_incommensurate_1990}. 
We expect that if we could probe such orders, parts of our phase diagram would be different.
Indeed, a VCA study~\cite{faye_interplay_2017} has found a charge density wave with a four-unit-cell period coexisting microscopically with superconductivity on the hole-doped side.
Although magnetism disappears at small doping on the hole-doped side even when superconductivity is absent, we cannot exclude that spiral order could persist to large hole dopings. 
This is one of the explanations offered~\cite{eberlein_fermi_2016} for the abrupt change in the Hall effect when one enters the pseudogap regime at low temperature and in magnetic fields sufficiently strong to destroy superconductivity. Collinear incommensurate magnetism, however, cannot explain this Hall data~\cite{charlebois_hall_2017}. 
It has been proposed that Seebeck measurements can tell apart the various phenomenological theories for this Hall data~\cite{verret_phenomenological_2017}. 

Comparing the two sets of band parameters, we observe that increasing second- and third-neighbour hopping reduces the regions of microscopic coexistence. 
This is understandable from the bare band structure that shows reduced nesting in that case~\cite{markiewicz_entropic_2017}, weakening antiferromagnetism. This effect is especially pronounced on the hole-doped side. 
The effect of $U$ on the amplitude of the triplet order hints that microscopic coexistence is more stable the closer the system is to the Mott transition, as the triplet order is stronger there.
Increasing $U$ also increases the domain of filling that supports microscopic coexistence on the electron-doped side.
The phase transitions are generally second order, except on the hole-doped side where the transition between antiferromagnetism and superconductivity is often first order. This reflects the weakness of antiferromagnetism away from half-filling on the hole-doped side.
\section{Conclusion}
Following the lead of Ref.~\cite{koch_sum_2008,liebsch_finite-temperature_2009, liebsch_temperature_2012}, we used symmetry and gauge-invariance considerations to propose the most general parametrization of the bath for an exact-diagonalization CDMFT solution of the Hubbard model with a 4-site cluster hybridized with 8 bath orbitals. 
The parametrization must be chosen according to the phases that are put in competition, here antiferromagnetism and $d$-wave superconductivity. 
A simpler parametrization gives qualitatively correct results only when antiferromagnetism and $d$-wave superconductivity do not coexist. 
We found phase diagrams that are much closer to observations than previous results found with simpler parametrizations. 

In particular, microscopic coexistence between antiferromagnetism and $d$-wave superconductivity is more robust for electron-doped compounds. 
For large $U$ and $|t'|$, the filling where antiferromagnetism ends in the absence of superconductivity is much larger than in the presence of superconductivity. 
 
Given the generality of the bath parametrization, our results are the most accurate that can be obtained with a finite bath and an exact diagonalization solver. They will be a useful guide for calculations that include an infinite bath but are performed with more resource-hungry continuous-time quantum Monte Carlo solvers.


\begin{acknowledgments}
This work has been supported by the Natural Sciences and Engineering Research Council of Canada (NSERC) under grants RGPIN-2014-04584 and RGPIN-2015-05598, the Canada First Research Excellence Fund, by the Research Chair in the Theory of Quantum Materials and by the Fonds de recherche du Qu\'ebec. Computing resources were provided by Compute Canada and Calcul Qu\'ebec.
\end{acknowledgments}
 
%

\end{document}